%% file: main.tex
\documentclass[a4paper]{article}

\usepackage[ninepoint]{itgspeech2025} %
\usepackage{times} %
\usepackage{standalone}
\usepackage[english]{babel} %
\usepackage[utf8]{inputenc} %
\usepackage[T1]{fontenc} %
\usepackage[sort&compress,numbers]{natbib} %
\usepackage{microtype}
\usepackage{amsmath,amssymb,amsfonts}
\usepackage{graphicx}
\usepackage[colorlinks=false,pdfborder={0 0 0}]{hyperref}
\usepackage[acronym,toc,shortcuts,nonumberlist]{glossaries}
\usepackage{xcolor}
\usepackage{xspace}
\usepackage{multirow}
\usepackage{subcaption}
\usepackage{hyperref}
\usepackage{cleveref}
\usepackage{siunitx}
\usepackage{pifont}
\usepackage[textsize=tiny,textwidth=1.4cm]{todonotes}

\glsdisablehyper
\input{commands}

\input{acronyms}
\input{figures/tikzstyle}

\title{Error Analysis in a Modular Meeting Transcription System}
\author{Peter Vieting\authorrefmark{*}, Simon Berger\authorrefmark{*}\authorrefmark{\ddagger}, Thilo von Neumann\authorrefmark{\mathsection}, Christoph Boeddeker\authorrefmark{\mathsection}, Ralf Schlüter\authorrefmark{*}\authorrefmark{\ddagger} and Reinhold Haeb-Umbach\authorrefmark{\mathsection}}
\address{%
\authorrefmark{*}Machine Learning and Human Language Technology Group, RWTH Aachen University, Germany\\Email: \texttt{\{vieting,berger,schlueter\}@hltpr.rwth-aachen.de}\\
\authorrefmark{\ddagger}AppTek GmbH, Germany\\
\authorrefmark{\mathsection}Paderborn University, Germany\\Email: \texttt{\{vonneumann,boeddeker,haeb\}@nt.upb.de}%
}

\begin{document}

\maketitle

\begin{abstract}
Meeting transcription is a field of high relevance and remarkable progress in recent years.
Still, challenges remain that limit its performance.
In this work, we extend a previously proposed framework for analyzing leakage in speech separation with proper sensitivity to temporal locality.
We show that there is significant leakage to the cross channel in areas where only the primary speaker is active.
At the same time, the results demonstrate that this does not affect the final performance much as these leaked parts are largely ignored by the \gls{vad}.
Furthermore, different segmentations are compared showing that advanced diarization approaches are able to reduce the gap to oracle segmentation by a third compared to a simple energy-based \gls{vad}.
We additionally reveal what factors contribute to the remaining difference.
The results represent state-of-the-art performance on LibriCSS among systems that train the recognition module on LibriSpeech data only.
\end{abstract}

\glsresetall

\section{Introduction}
Meeting transcription is a task of increasing importance as it is key to enable a diverse set of applications.
Different approaches have been proposed for meeting transcription that can be grouped into modular (e.g. \cite{chen2021css_conformer,berger2023interspeech}) and end-to-end systems (e.g. \cite{sklyar2022multiturn,kanda2022tsot}).
While modular systems consist of cascaded submodules and are more interpretable, end-to-end approaches have the advantage of taking only one single global decision.
Despite impressive progress in recent years \cite{chen2020libricss,kanda2022tsot,wang2023tfgridnet_transactions,boeddeker2024tssep}, meeting transcription is still challenging \cite{yu2022m2met}.
This is due to different factors such as a setting with far-field recordings and noisy acoustic conditions, the requirement of accurate speaker attribution depending on the application and the fact that separation of overlapping speech is still facing challenges such as leakage.
In this work, we investigate components and factors that contribute to errors of a modular system.

The impact of speech enhancement errors on \gls{asr} has been studied in \cite{iwamoto2022artifacts,araki2023artifacts}.
Similarly, separation artifacts can affect the downstream performance \cite{cordlandwehr2022monaural}.
However, the thorough analysis of leakage in speech separation systems and its impact on subsequent \gls{asr} has received little attention.
With leakage, we refer to situations where a separation output channel should be silent but instead contains audio that was either partially moved or copied there from the other channel.
Common metrics that are computed purely on the transcription using Levenshtein distances (i.e., \gls{wer} and its variants) may blur leakage effects as they do not guarantee temporal locality.
On the other hand, signal-level metrics may struggle to handle silence appropriately and may penalize properties that do not degrade the final \gls{wer} \cite{iwamoto2022artifacts}.
These challenges can be addressed using Hamming distances of frame-wise word-level alignments \cite{wessel2001errorminimization, wessel2001confidencemeasures}.

We recently proposed a framework to detect leakage in a modular meeting transcription system using frame-wise word-level alignments \cite{vieting2024mixture_encoder}.
This framework is applied and extended here to have a more detailed look at different types of leakage that may occur.
In particular, we enable the measurement of leakage in situations that were not covered in our previous analysis.
Furthermore, we investigate which specific error types contribute to the significant degradation caused by an imperfect segmentation that was observed in \cite{vieting2024mixture_encoder}.

The extended framework is applied to analyze a strong modular meeting transcription system that follows the \gls{css} idea \cite{yoshioka2018CSS,chen2020libricss} and utilizes TF-GridNet \cite{wang2023tfgridnet_transactions, wang2023tfgridnet_icassp} for speech separation.
We extend our system with a diarization module that utilizes \gls{asr} transcriptions to refine the segmentation \cite{vonneumann2024diarization} and demonstrate that this improves performance even for measures that do not penalize speaker attribution errors.
In addition to the original diarization result, we replace the Whisper \gls{asr} \cite{radford2023whisper} with our own system in the diarization pipeline to remove external dependencies that apply large models and require extensive amounts of data and compute.
The results demonstrate a competitive performance on LibriCSS and outperform other systems that only train on LibriSpeech data.

The main contributions of this work are
\begin{itemize}
    \item the extension of a leakage analysis framework that allows measuring significant leakage to the cross channel in areas where only the primary speaker is active,
    \item a detailed breakdown of segmentation error types and their contribution to the gap to the oracle segmentation performance,
    \item state-of-the-art results for single-microphone meeting transcription on the LibriCSS task among systems that train the \gls{asr} module on LibriSpeech data only.
\end{itemize}

\section{Meeting Transcription}
\label{sec:pipeline}
Speech separation, recognition and diarization modules can be composed in different orders to form a meeting transcription pipeline \cite{raj2021integration}.
This section describes the pipeline used in this work, which is similar to \cite{vonneumann2024diarization}.
This pipeline can shortly be described as \gls{css} followed by segmentation, \gls{asr}, and diarization and is thus referred to as \glsunset{cssad}\gls{cssad}.
\Cref{fig:pipeline} depicts an overview.

\subsection{Continuous Speech Separation}
\label{sec:css}
The \gls{css} idea \cite{yoshioka2018CSS,chen2020libricss} allows speaker separation for an arbitrary number of speakers.
A separation module separates the observed signal into two overlap-free signals within a small window.
This is possible under the assumption that the window size is small enough to ensure that at maximum two speakers are active within a single window.
A given speaker might be assigned to different output channels by the separator when shifting the window.
This is referred to as the permutation problem.
It is tackled by enabling a small overlap between neighboring windows and choosing the permutation of adjacent segments that results in a minimal \gls{mse} on the overlapping parts.

\subsection{Segmentation}
\label{sec:segmentation}

\input{figures/pipeline_visualization}

The separated audio channels are then segmented with a \gls{vad} module that extracts the speech regions.
We use an energy-based \gls{vad} that uses the ratio of the energies across both separated channels to identify active speech.
Using both channels is more robust against leakage than the classical approach of only looking at one channel to find the speech activity.

Note that in principle, the two separation output channels are equal in the sense that there is no notion of primary and cross channel.
However, as soon as we consider a given segmentation, the speech segments are not positioned at the same times in both channels.
For a given segment, we refer to the channel that contains the separated speech of that segment as the primary channel (e.g. the channel with index 1).
The cross channel (in that case the channel with index 2) may contain silence or another speaker that overlapped in the original mixture observation.
This is important to keep in mind for the analysis in \Cref{sec:leakage_analysis}.

\subsection{Recognition}
The segments containing separated single-speaker speech can be recognized using \gls{asr}.
It outputs the transcription of what has been spoken in the given segment.
While different architectures are conceivable, we use a hybrid hidden Markov model as in \cite{vieting2024mixture_encoder}.
It has the advantage of providing accurate word-level timestamps.

\subsection{Diarization}
\label{sec:diarization}
We use the \gls{asr}-supported diarization approach from \cite{vonneumann2024diarization} with slight modifications.
It utilizes the word-level timestamps from the \gls{asr} to further detect speaker changes that were undetected by \gls{vad} alone.
The segmentation returned by the diarization module can thus be different from the segmentation of the \gls{vad}.
Note that it is possible to rerun the \gls{asr} on the refined segmentation of the \gls{cssad} output and this is the approach we adopt in this work.

\section{Methods}
\subsection{Leakage Analysis}
\label{sec:leakage_analysis}

In \cite{vieting2024mixture_encoder}, we measured the effect of leakage in order to understand how separation errors influence the \gls{asr} performance.
For this analysis, we used \glspl{cr}, which measure the fraction of frames where the word-level alignments of both channels match.
These matches are an indication of a possible spill-over of one channel into the other.
We distinguish between \gls{wcr} on single-best hypotheses and \gls{gcr} on lattices in analogy to word and graph error rates.
Note that the \gls{gcr} is optimistic because it checks per time frame whether the word labels match for any arc in the lattice irrespective if the considered arcs compose a valid path.
An example for the \gls{gcr} computation is depicted in \Cref{fig:fer}.

As explained in \Cref{sec:segmentation}, the notion of a primary and a cross channel arises when considering a segmentation of the data.
For a given segment, the primary channel is the channel containing the speech (e.g. channel 1) while the cross channel (in that example channel 2) may contain parts of another speech segment or silence.
We still consider all segments from all channels.
As a consequence, channel 2 in this example is considered as the primary channel for segments located on that channel.

We showed that the cross channel is well suppressed and that words leaked from cross-talkers into the primary channel hardly play a role in the primary channel's search space \cite{vieting2024mixture_encoder}.
This means that for a given segment, the \gls{asr} hypotheses rarely contain words from the cross channel's forced aligned ground-truth transcription.
However, the motivation was to study only the direct impact on the recognition performance to assess the feasibility of utilizing cross-speaker transcriptions in sequence discriminative training.
Thus, only leakage from the cross-talker onto the primary channel was analyzed within the boundaries of the oracle segments.
An example is depicted in \Cref{fig:analysis_directions} on the left for the green segment, where possibly leaked words from the red speaker into the green segment ("dolor", "sit" or "amet") would have been measured.

\input{figures/fer_visualization.tex}

\input{figures/macro_cir_viz}

In this work, we extend the analysis by measuring leakage from the primary channel to the cross channel, which includes regions where no speech is active on the cross channel.
These regions are typically outside the segments and were therefore previously ignored.
Furthermore, leakage from the primary channel to the cross channel could not be measured before because only forced alignments on the cross channel were considered that could not contain leaked words which are not in the transcription.
An example is given by the red transcription that does not contain the green leak in \Cref{fig:analysis_directions}.
Now, we also compare forced alignments of the primary speaker to hypotheses on the cross channel as shown in \Cref{fig:analysis_directions} on the right for the blue segment.
When recognizing the cross channel signal, the blue leak is then likely to be present in the resulting hypothesis.
The \gls{gcr} computation for this example is shown by \Cref{fig:fer}.

\input{tables/results_diarizations.tex}
Beyond the above extension, a few general improvements are applied.
Previously, we used forced alignments obtained on the separated LibriCSS signals.
However, this may result in problematic alignments where the separation creates artifacts, which are the most interesting positions, in fact.
To mitigate potential biases, we now use forced alignments obtained on the clean LibriSpeech signals and synchronize them to LibriCSS.
Furthermore, the \gls{smbr} fine-tuned model with the best final performance is used to generate the hypotheses instead of the \gls{fce} model.
Finally, we now apply the same regular search settings for lattices as for the 1-best case instead of settings targeted for \gls{smbr} training.

\subsection{Segmentation Analysis}
Motivated by the large performance gap of our system with \gls{vad} segmentation to the oracle segmentation in \cite{vieting2024mixture_encoder}, we aim to study the errors that occur during segmentation.
We used the visualization tool from the MeetEval toolkit \cite{neumann2023meeteval} to find typical problems.
For each error type, we define a heuristic that exploits oracle information to eliminate these errors from the segmentation.
The \gls{wer} obtained on the refined segmentation is then compared to the \gls{wer} on the original segmentation to assess the impact of each error type.

\vspace{-1.0ex}
\section{Experimental Setup}
\vspace{-0.5ex}
\subsection{Data}
We evaluate our models on the LibriCSS dataset \cite{chen2020libricss}.
It features re-recordings of utterances from the LibriSpeech \cite{panayotov2015librispeech} \textit{test-clean} set in meeting rooms with varying degrees of overlap.
LibriCSS is well suited as an evaluation task for this work because of its meeting-like structure, real-world acoustic conditions and because of its widespread use in the research field \cite{chen2020libricss,kanda2022tsot,boeddeker2024tssep}.

In this work, we address single-microphone meeting transcription.
We therefore only use the first microphone of the LibriCSS data.
Furthermore, \textit{Session0} is used as a dev set to tune hyperparameters as suggested in \cite{raj2021integration}, and we report the results on the remaining sessions.

Since LibriCSS only provides data for evaluation, we use the LibriSpeech signals to simulate spatialized and mixed training data similar to \cite{drude2019smswsj}.
The speech separators are trained on this data.
The \gls{asr} model is first trained on clean LibriSpeech and then fine-tuned on the signals that are obtained by applying the separator to the simulated data.
For more details, see \cite{vieting2024mixture_encoder}.
The LibriSpeech text corpus is used to train the \glspl{lm} \cite{irie2019trafolm}.

\subsection{Meeting Transcription}
\label{sec:system_description}
In this work, we use the meeting transcription system from \cite{vieting2024mixture_encoder} and extend it with a diarization module.
The pipeline is outlined in \Cref{sec:pipeline}.

TF-GridNet \cite{wang2023tfgridnet_transactions, wang2023tfgridnet_icassp} is used to separate the observed signal into two overlap-free signals within a sliding window with a size of \SI{4}{\second} and a shift of \SI{3}{\second} as described in \Cref{sec:css}.
For segmentation, the baseline is a simple energy-based \gls{vad} \cite{vieting2024mixture_encoder}.
In addition, we evaluate the refined segmentation obtained by the different diarization systems.

\Gls{asr} is performed using a hybrid hidden Markov model.
The neural encoder consists of 12 conformer blocks \cite{gulati2020conformer} and has a total of 87M parameters.
We use the \gls{smbr} fine-tuned model with baseline encoder from \cite{vieting2024mixture_encoder}.
Further details are outlined in \cite{vieting2024mixture_encoder}.
During recognition, we use the official LibriSpeech 4gram \gls{lm} as well as a neural \gls{trafolm}.

The \gls{asr}-supported diarization pipeline builds on top of the word-level timestamps from \gls{asr} \cite{vonneumann2024diarization}.
For every word boundary, one speaker embedding vector is computed each for the left and right context of \SI{3}{\second} around the boundary.
If the similarity between the two is below a threshold and the lowest among the context of \SI{4}{\second}, it is considered as a speaker change, and the segment is split.
Afterwards, one speaker embedding vector is extracted for every segment and a k-means clustering is applied to obtain speaker labels.
Note that the initial \gls{vad} hyperparameters are selected differently here to create shorter segments and avoid segments that contain speaker changes.

Finally, we pass the refined segments again to \gls{asr} to obtain more accurate transcriptions.
Here, we additionally merge subsequent segments if the diarization assigns them to the same speaker and the pause between the segments is not longer than \SI{3}{\second} to have more context within the segments if possible.

\section{Results}
\Cref{table:results_diarizations} presents the \gls{cpwer} of our meeting transcription system on the LibriCSS task.
Our results are competitive with existing works and outperform the systems in \cite{vonneumann2024diarization,boeddeker2024interspeech} that both use a large pre-trained WavLM\footnote{\url{https://huggingface.co/espnet/simpleoier_librispeech_asr_train_asr_conformer7_wavlm_large_raw_en_bpe5000_sp}} \cite{chen2022wavlm} model for \gls{asr}.
The \gls{cpwer} for both of our diarizations is identical and constitutes a new state-of-the-art performance for systems that only use LibriSpeech data for \gls{asr} training.
Solely the DCF-DS system in combination with WavLM obtains a better single-microphone \gls{cpwer} on LibriCSS.
Notably, our results are achieved without external dependencies that apply large models and require extensive amounts of data or compute.
Only the speaker embedding extractor was trained on data other than LibriSpeech, namely VoxCeleb.

To compare our systems to the baseline \gls{vad} segmentation in \cite{vieting2024mixture_encoder}, \Cref{table:results_segmentations} reports the \gls{orcwer} \cite{sklyar2022multiturn}.
Unlike the \gls{cpwer}, it does not account for speaker attribution errors and is therefore generally lower.
The oracle segmentation is obtained by using the boundaries of the original LibriSpeech utterance provided by the LibriCSS annotation and selecting the separated channel with minimum \gls{sdr} \cite{vincent2006performance} to the clean audio.
\Cref{table:results_segmentations} compares the oracle and \gls{vad} results to the refined segmentations from diarization based on either Whisper's or our transcription are tested.
In specific, Whisper \cite{radford2023whisper} was deployed in the "large-v2" configuration.
Note that the hyperparameters for the preceding \gls{vad} are tuned individually for the latter two cases.
We can observe clear improvements compared to \cite{vieting2024mixture_encoder}, closing around a third of the previous gap to the oracle performance.
\input{tables/results_segmentations.tex}

\subsection{Leakage Analysis}
\label{sec:results_leakage}
\Cref{table:results_leakage} presents the results of the leakage analysis.
We report \glspl{cr} which describe the fraction of frames in which both sequences contain the same word.
The first line (spoken vs. spoken) represents the natural coincidence, i.e., how often both speakers really uttered the same word at the same time.
Leakage from the cross channel to the primary channel is considered in the next part of the table (1-best/lattice vs. spoken), similarly to \cite{vieting2024mixture_encoder}.
The results do not deviate much despite our updates in the analysis.
The \glspl{cr} for the primary channel hypotheses (1-best and lattice) with the cross channel ground truth (spoken) are higher than the natural coincidence, but this is mainly caused by silence.
We observe higher \glspl{cr} for words in regions with two active speakers and for silence generally compared to \cite{vieting2024mixture_encoder}.
However, the overall conclusion that the cross speaker is well suppressed and does not have a major impact on the primary channel's search space still holds.

Finally, leakage from the primary channel to the cross channel is addressed (spoken vs. 1-best/lattice).
Significant leakage can be observed for this direction in areas where only one speaker is active.
This can likely be explained by the nature of the segmentation.
In a given segment, there are few positions where the cross speaker is active and the primary speaker is not because the segment is targeted to the primary speaker.
In contrast, there are many frames where only the primary speaker is active and therefore more chances to create leakage from the primary channel to the cross channel.

This could be a hint why there is such a significant performance gap to the oracle segmentation.
If leakage of this type occurs, the \gls{vad} could create a segment for the leaked speech which would be transcribed and cause edits in the \gls{wer}.
The oracle segmentation automatically discards these leaks.
If this is really a major cause of recognition errors, will be investigated in the next section.

\subsection{Segmentation Analysis}
By manual inspection of the MeetEval visualizations, we identified the following segmentation error types:
\begin{enumerate}
    \item The cross channel should be silent but audio from the primary channel leaks through, either by partially moving or by copying the audio to the cross channel ("leakage"). This is the error type analyzed in the previous section.
    \item Some parts of segments are missing, removing relevant speech contents ("missing").
    \item The segmentation creates long segments that merge several oracle segments into one ("merges").
    \item Even in the case of clear correspondence of \gls{vad} and oracle segments, the boundary times typically deviate slightly ("boundaries").
\end{enumerate}

\input{tables/results_leakage.tex}

\input{tables/results_segmentations_fixes.tex}

\Cref{table:results_segmentations_fixes} shows the impact of running \gls{asr} on a segmentation that eliminates these different error types using oracle information.
Unlike the results in \Cref{sec:results_leakage} might suggest, removing leaked segments does not result in a major improvement.
Our best \gls{orcwer} even remains unchanged.
Similarly, splitting of segments that consist of multiple oracle segments and adjusting the boundary times does not have a significant effect on the performance either.
In this case, this can be considered as expected.
Splitting segments according to the oracle segmentation mainly leads to a smaller context size for the \gls{lm} which might hurt or not depending on how related the individual segments are.
Note that in some cases, the subsequent segments are subsequent segments in LibriSpeech \textit{test-clean} suggesting that the longer context could even be useful.
Adapting the boundaries should mainly only affect silence at the beginning and end of segments which is not expected to have a strong impact on the recognized hypothesis.

However, adding missing segments significantly improves the \gls{orcwer}.
This error type accounts for more than half of the performance gap to the oracle segmentation. %
In combination, fixing the observed error types allows closing around 90\% of the gap for the Whisper-informed diarization and around 75\% for our diarization. %
An absolute difference of only 0.1\% or 0.2\% respectively remains.
Future work might therefore concern improved \gls{vad} to address these errors.

\section{Conclusion}
This work studies remaining errors in a strong modular meeting transcription system.
For this, we extend an existing leakage analysis framework with proper sensitivity to temporal locality and show that significant leakage to the cross channel can be measured in regions where only the primary speaker is active.
After identifying typical segmentation error types, we evaluate their effect on the performance gap to the oracle segmentation.
The results show that missing speech segments are the main contributor and that the identified leakage is not a major problem.
By adding advanced diarization systems, we close around a third of the gap to the oracle segmentation and achieve state-of-the-art single-microphone results on the LibriCSS task among systems that only use LibriSpeech data for \gls{asr} training.

\section{Acknowledgement}
This research was partially funded by the Deutsche Forschungsgemeinschaft (DFG, German Research Foundation) under project No. 448568305.
Computational Resources were provided by BMBF/ NHR/ PC2.

\bibliographystyle{IEEEtran}
\bibliography{refs}

\end{document}

%% file: commands.tex
\newcommand*\nestedglsentry[1]{%
  \protect\ifglsused{#1}{%
    \glsentryshort{#1}%
  }{%
    \glsentrylong{#1}%
  }%
}
\newcommand{\authorrefmark}[1]{\textsuperscript{\footnotesize\ensuremath{#1}}}
\newcommand{\libricss}{LibriCSS\xspace}

%% file: acronyms.tex
\newacronym{aed}{AED}{attention-based encoder decoder}
\newacronym{am}{AM}{acoustic model}
\newacronym{asr}{ASR}{automatic speech recognition}
\newacronym[longplural={bi-directional long-short term memories}]{blstm}{BLSTM}{bi-directional long-short term memory}
\newacronym{ce}{CE}{cross entropy}
\newacronym{cr}{CR}{coincidence rate}
\newacronym{cnn}{CNN}{convolutional neural network}
\newacronym{css}{CSS}{continuous speech separation}
\newacronym{cssad}{CSS-AD}{\nestedglsentry{css} followed by \nestedglsentry{asr} and diarization}
\newacronym{ctc}{CTC}{connectionist temporal classification}
\newacronym{fer}{FER}{frame error rate}
\newacronym{fir}{FIR}{finite impulse response}
\newacronym{fce}{f-CE}{frame-wise cross-entropy}
\newacronym{gcr}{GCR}{graph \nestedglsentry{cr}}
\newacronym{ger}{GER}{graph error rate}
\newacronym{gpu}{GPU}{graphics processing unit}
\newacronym{her}{HER}{Hamming error rate}
\newacronym{lm}{LM}{language model}
\newacronym[longplural={long-short term memories}]{lstm}{LSTM}{long-short term memory}
\newacronym{mse}{MSE}{mean squared error}
\newacronym{nn}{NN}{neural network}
\newacronym{sa-sdr}{SA-SDR}{source-aggregated signal-to-distortion ratio}
\newacronym{sdr}{SDR}{signal-to-distortion ratio}
\newacronym{snr}{SNR}{signal to noise ratio}
\newacronym{smbr}{sMBR}{state-level minimum Bayes risk}
\newacronym{stft}{STFT}{short time Fourier transform}
\newacronym{trafolm}{Trafo LM}{transformer-based \nestedglsentry{lm}}
\newacronym{vad}{VAD}{voice activity detection}
\newacronym{wcr}{WCR}{word \nestedglsentry{cr}}
\newacronym{wer}{WER}{word error rate}
\newacronym{cpwer}{cpWER}{concatenated minimum-permutation \nestedglsentry{wer}}
\newacronym{orcwer}{ORC WER}{optimal reference combination \nestedglsentry{wer}}
\newacronym{werr}{WERR}{word error rate reduction}

%% file: figures/tikzstyle.tex
\usepackage{tikz}

\usepackage{tikzsymbols}

\usepackage[outline]{contour}
\contourlength{1pt}

\usetikzlibrary{calc}
\usetikzlibrary{positioning}
\usetikzlibrary{fit}
\usetikzlibrary{arrows}
\usetikzlibrary{decorations.pathreplacing}
\usetikzlibrary{arrows.meta}
\usetikzlibrary{backgrounds}

\tikzset{>=stealth}
\tikzstyle{block}=[
    draw,
    text depth=0pt,
    thick, 
    rectangle,
    text centered,
    minimum height=3.4ex,
    fill=white,
] %

\definecolor{muted0}{HTML}{4878d0}
\definecolor{muted1}{HTML}{ee854a}
\definecolor{muted2}{HTML}{6acc64}
\definecolor{muted3}{HTML}{d65f5f}
\definecolor{muted4}{HTML}{956cb4}
\definecolor{muted5}{HTML}{8c613c}
\definecolor{muted6}{HTML}{dc7ec0}
\definecolor{muted7}{HTML}{797979}
\definecolor{muted8}{HTML}{d5bb67}
\definecolor{muted9}{HTML}{82c6e2}

\definecolor{peterBlue}{HTML}{032F74}
\definecolor{peterOrange}{HTML}{B15027}

\tikzset{%
    do path picture/.style={%
        path picture={%
            \pgfpointdiff{\pgfpointanchor{path picture bounding box}{south west}}%
            {\pgfpointanchor{path picture bounding box}{north east}}%
            \pgfgetlastxy\x\y%
            \tikzset{x=\x/2,y=\y/2}%
            #1
        }
    },
}

%% file: figures/pipeline_visualization.tex
\begin{figure*}[t]

\newcommand{\wavA}{\includegraphics[height=1.1em,width=1.5em]{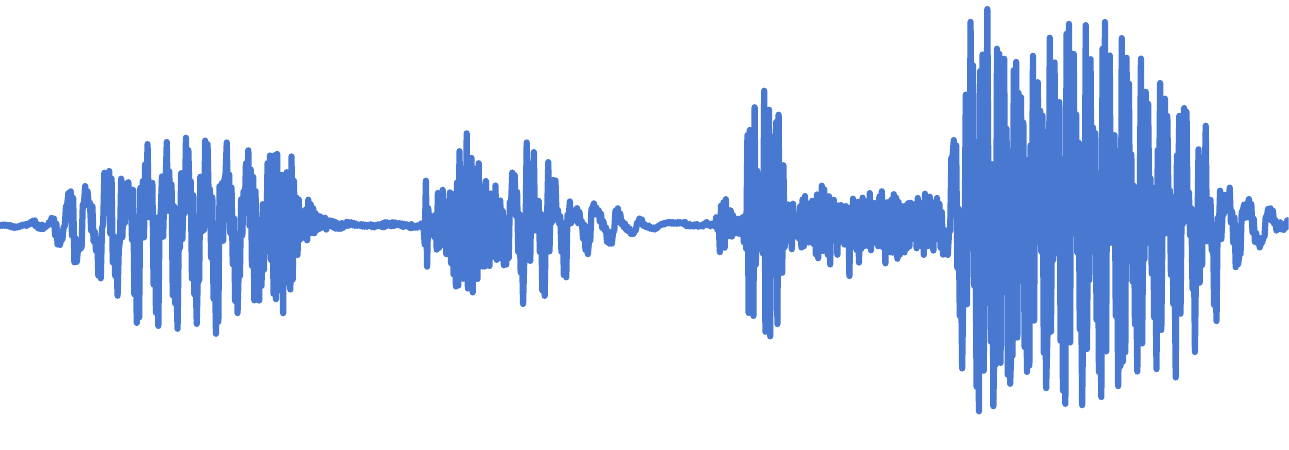}}
\newcommand{\wavB}{\includegraphics[height=0.9em,width=.5em,trim={0 0 12.2cm 0}, clip]{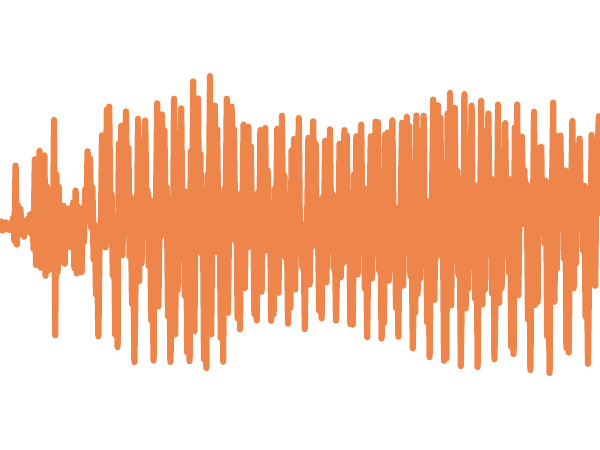}}
\newcommand{\wavC}{\includegraphics[height=1.1em,width=.5em]{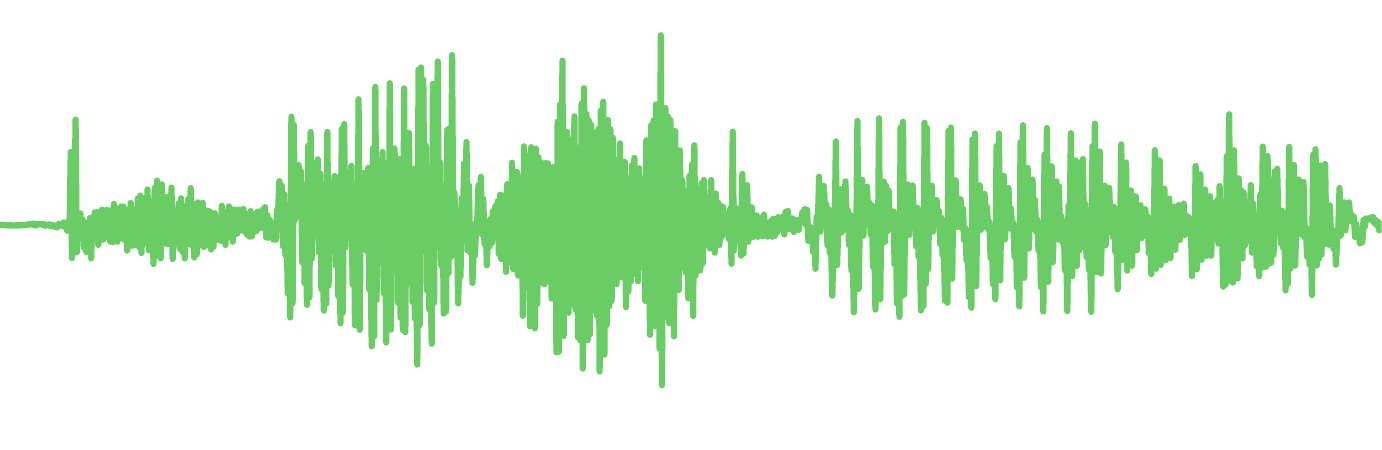}}
\newcommand{\wavD}{\includegraphics[height=1.1em,width=.6em,trim={10.2cm 0 0 0}, clip]{figures/wav/wav_2_C1.pdf}}
\newcommand{\wavBshift}{-0.4em}
\newcommand{\wavCshift}{0.2em}
\newcommand{\wavDshift}{0.15em}
\newcommand{\channelShift}{-0.45}
\centering%
\begin{tikzpicture}[
    x=1em, y=6ex, 
    scale=1, 
    line width=0.1em,
    mywavstyle/.style={inner xsep=0, inner ysep=0, fill opacity=0.3, outer sep=0, anchor=west,minimum height=1.1em},
    box/.style={draw,rounded corners=1,thin, inner sep = .5em},
    arrow/.style={draw,->,thin,rounded corners=1},
]
 
\node[mywavstyle, inner sep=0, inner xsep=0] (wav1) at (0, 0) {\wavA};
\node[mywavstyle, anchor=center] (wav2) at  ($(wav1.west)+(\wavBshift,0)$) {\wavB};
\node[mywavstyle, anchor=center] (wav3) at  ($(wav1.east)+(\wavCshift,0)$) {\wavC};
\node[mywavstyle, anchor=center] (wav4) at  ($(wav1.west)+(\wavDshift,0)$) {\wavD};
\node[mywavstyle, anchor=center, opacity=0.5] (wav3) at  ($(wav1.east)+(\wavCshift,0)$) {\wavC};
\node[mywavstyle, anchor=center, opacity=0.5] (wav2) at  ($(wav1.west)+(\wavBshift,0)$) {\wavB};
\node[mywavstyle, inner sep=0, inner xsep=0, opacity=0.5] (wav1) at (0, 0) {\wavA};

\node[anchor=west,box] at ($(wav1.east) + (1em,0)$) (css) {CSS};
\node[anchor=west,box] at ($(css.east) + (5.25em,0)$) (vad) {VAD Segmentation};
\node[anchor=west,box] at ($(vad.east) + (5.25em,0)$) (asr) {ASR};
\node[anchor=west,box] at ($(asr.east) + (5.25em,0)$) (dia) {Diarization};
\node[anchor=west,box] at ($(dia.east) + (5.25em,0)$) (asr2) {ASR};

\draw[arrow] ($(wav1.east) + (.5em,0)$) -- (css);
\draw[arrow] (css) -- coordinate[yshift=0,xshift=-.125em] (coord-pict-sep) (vad);
\draw[arrow] (vad) -- coordinate[xshift=-.125em] (coord-pict-vad) (asr);
\draw[arrow] (asr) -- coordinate[xshift=-.125em] (coord-pict-asr) (dia);
\draw[arrow] (dia.east) -- coordinate[xshift=-.125em] (coord-pict-dia) (asr2.west);
\draw[arrow] (asr2.east) -- ++(1em,0) coordinate[xshift=2.25em] (coord-pict-asr2);

\node[draw, gray, very thin, fill=white,minimum height=2.5em, minimum width=4.25em,rounded corners=2] at (coord-pict-sep) {};
\begin{scope}
\clip[rounded corners=2] ($(coord-pict-sep) - (2.125em,1.25em)$) rectangle ++(4.25em, 2.5em);
\draw[very thin, black] 
    ($(coord-pict-sep) + (0,-.5*\channelShift) + (-.625em,0)$) -- ($(coord-pict-sep) + (0,-.5*\channelShift) + (3em,0)$);
\draw[very thin, black] ($(coord-pict-sep) + (0,.5*\channelShift) + (-.625em,0)$) -- ($(coord-pict-sep) + (0,.5*\channelShift) + (3em,0)$);
\node[mywavstyle, anchor=center] (pict-sep-wav1) at ($(coord-pict-sep) + (.865em,-.5*\channelShift)$) {\wavA};
\node[mywavstyle, anchor=center] (pict-sep-wav2) at  ($(pict-sep-wav1.west)+(\wavBshift,0)$) {\wavB};
\node[mywavstyle, anchor=center] (pict-sep-wav3) at  ($(pict-sep-wav1.east)+(\wavCshift,\channelShift)$) {\wavC};
\node[mywavstyle, anchor=center] (pict-sep-wav4) at  ($(pict-sep-wav1.west)+(\wavDshift,\channelShift)$) {\wavD};
\node[anchor=east] at ($(pict-sep-wav1.west) + (-.4,0)$) (sep-lbl1) {\tiny Chn1:};
\node[anchor=east] at (pict-sep-wav3.west-|sep-lbl1.east) {\tiny Chn2:};
\end{scope}

\node[draw, gray, very thin, fill=white,minimum height=2.5em, minimum width=4.25em,rounded corners=2] at (coord-pict-vad) {};
\begin{scope}
\clip[rounded corners=2] ($(coord-pict-vad) - (2.125em,1.25em)$) rectangle ++(4.25em, 2.5em);
\node[mywavstyle, fill=muted0, anchor=center] (pict-vad-wav1) at ($(coord-pict-vad) + (0.865,-.5*\channelShift)$) {\wavA};
\node[mywavstyle, anchor=center, fill=muted1] (pict-vad-wav2) at  ($(pict-vad-wav1.west)+(\wavBshift,0)$) {\wavB};
\node[mywavstyle, anchor=center, fill=muted2] (pict-vad-wav3) at  ($(pict-vad-wav1.east)+(\wavCshift,\channelShift)$) {\wavC};
\node[mywavstyle, anchor=center,fill=muted1] (pict-vad-wav4) at  ($(pict-vad-wav1.west)+(\wavDshift,\channelShift)$) {\wavD};
\node[anchor=east] at ($(pict-vad-wav1.west) + (-.4,0)$) (vad-lbl1) {\tiny Chn1:};
\node[anchor=east] at (pict-vad-wav3.west-|vad-lbl1.east) {\tiny Chn2:};
\end{scope}

\node[draw, gray, very thin, fill=white,minimum height=2.5em, minimum width=4.25em,rounded corners=2] at (coord-pict-asr) {};
\begin{scope}
\clip[rounded corners=2] ($(coord-pict-asr) - (2.125em,1.25em)$) rectangle ++(4.25em, 2.5em);
\node[mywavstyle, anchor=center, fill=muted0] (pict-asr-t1) at ($(coord-pict-asr) + (0.865,-.5*\channelShift)$) {\wavA};
\node[color=black, color=muted0, inner sep=0.05em] at (pict-asr-t1) {\tiny \contour{white}{hello}};
\node[mywavstyle, anchor=center, fill=muted1] (pict-asr-t2) at  ($(pict-asr-t1.west)+(\wavBshift,0)$) {\wavB};
\node[color=muted1] at (pict-asr-t2) {\tiny \contour{white}{I}};
\node[mywavstyle, anchor=center, fill=muted2] (pict-asr-t3) at  ($(pict-asr-t1.east)+(\wavCshift,\channelShift)$) {\wavC};
\node[color=muted2] at (pict-asr-t3) {\tiny \contour{white}{a}};
\node[mywavstyle, anchor=center,fill=muted1] (pict-asr-t4) at  ($(pict-asr-t1.west)+(\wavDshift,\channelShift)$) {\wavD};
\node[color=muted1] at (pict-asr-t4) {\tiny \contour{white}{n}};
\node[anchor=east] at ($(pict-asr-t1.west) + (-.4,0)$) (asr-lbl1) {\tiny Chn1:};
\node[anchor=east] at (pict-asr-t3.west-|asr-lbl1.east) {\tiny Chn2:};
\end{scope}

\node[draw, gray, very thin, fill=white,minimum height=2.5em, minimum width=4.25em,rounded corners=2] at (coord-pict-dia) {};
\begin{scope}
\clip[rounded corners=2] ($(coord-pict-dia) - (2.25em,1.25em)$) rectangle ++(4.5em, 2.5em);
\node[mywavstyle,  fill=muted0, anchor=center] (pict-dia-t1) at ($(coord-pict-dia) + (0.865,-.5*\channelShift)$) {\wavA};
\node[color=muted0] at (pict-dia-t1) {\tiny a.b};
\node[color=black, color=muted0, inner sep=0.05em] at (pict-dia-t1) {\tiny \contour{white}{hello}};
\node[mywavstyle, anchor=center, fill=muted1] (pict-dia-t2) at  ($(pict-dia-t1.west)+(\wavBshift,.5*\channelShift)$) {\wavB};
\node[color=muted1] at (pict-dia-t2) {\tiny \contour{white}{I}};
\node[color=muted1] at (pict-dia-t2) {\tiny \contour{white}{I}};
\node[mywavstyle, anchor=center, fill=muted2] (pict-dia-t3) at  ($(pict-dia-t1.east)+(\wavCshift,\channelShift)$) {\wavC};
\node[color=muted2] at (pict-dia-t3) {\tiny \contour{white}{a}};
\node[mywavstyle, anchor=center,fill=muted1] (pict-dia-t4) at  ($(pict-dia-t1.west)+(\wavDshift,.5*\channelShift)$) {\wavD};
\node[color=muted1] at (pict-dia-t4) {\tiny \contour{white}{n}};
\node[anchor=east] at ($(pict-dia-t1.west) + (-.4,0)$) (dia-lbl1) {\tiny Spk1:};
\node[anchor=east] at (pict-dia-t2.west-|dia-lbl1.east) {\tiny Spk2:};
\node[anchor=east] at (pict-dia-t3.west-|dia-lbl1.east) {\tiny Spk3:};
\end{scope}

\node[draw, gray, very thin, fill=white,minimum height=2.5em, minimum width=4.25em,rounded corners=2] at (coord-pict-asr2) {};
\begin{scope}
\clip[rounded corners=2] ($(coord-pict-asr2) - (2.25em,1.25em)$) rectangle ++(4.5em, 2.5em);
\node[mywavstyle,  fill=muted0, anchor=center] (pict-asr2-t1) at ($(coord-pict-asr2) + (0.865,-.5*\channelShift)$) {\wavA};
\node[color=muted0] at (pict-asr2-t1) {\tiny a.b};
\node[color=black, color=muted0, inner sep=0.05em] at (pict-asr2-t1) {\tiny \contour{white}{hello}};
\node[mywavstyle, anchor=center, fill=muted1] (pict-asr2-t2) at  ($(pict-asr2-t1.west)+(\wavBshift,.5*\channelShift)$) {\wavB};
\node[color=muted1] at (pict-asr2-t2) {\tiny \contour{white}{I}};
\node[mywavstyle, anchor=center, fill=muted2] (pict-asr2-t3) at  ($(pict-asr2-t1.east)+(\wavCshift,\channelShift)$) {\wavC};
\node[color=muted2] at (pict-asr2-t3) {\tiny \contour{white}{a}};
\node[mywavstyle, anchor=center,fill=muted1] (pict-asr2-t4) at  ($(pict-asr2-t1.west)+(\wavDshift,.5*\channelShift)$) {\wavD};
\node[color=muted1] at ($(pict-asr2-t2.west)!0.5!(pict-asr2-t4.east)$) {\tiny \contour{white}{I'm}};
\node[anchor=east] at ($(pict-asr2-t1.west) + (-.4,0)$) (dia-lbl1) {\tiny Spk1:};
\node[anchor=east] at (pict-asr2-t2.west-|dia-lbl1.east) {\tiny Spk2:};
\node[anchor=east] at (pict-asr2-t3.west-|dia-lbl1.east) {\tiny Spk3:};
\end{scope}

\end{tikzpicture}
\vspace{-1.0ex}
\caption{Meeting transcription pipeline. A \acrfull{css} system separates overlapping speech of multiple speakers into two overlap-free channels. A \acrfull{vad} identifies regions with speech activity. \Acrfull{asr} transcribes each detected segment individually. The diarization assigns speaker labels and refines the segmentation using \gls{asr} information.
\Gls{asr} can again transcribe the resulting segments.
Different colors represent different ground-truth speakers.}
\label{fig:pipeline}
\vspace{-1.5ex}
\end{figure*}

%% file: figures/fer_visualization.tex
\begin{figure}[tb]
    \includegraphics[width=0.95\columnwidth]{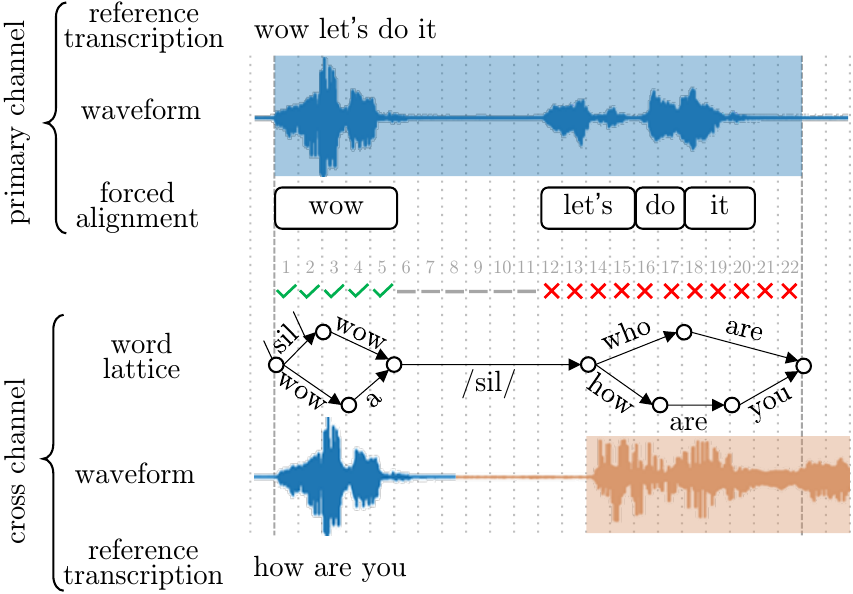}
    \caption{
        Visualization of \gls{gcr} computation measured between the primary channel forced alignment and the cross channel word lattice.
        We indicate whether the word labels match for any arc ({\color{green}\ding{52}}), mismatch ({\color{red}\ding{54}}) or both channels' forced alignments contain silence ({\color{gray}\textbf{\Large{--}}}).
        The example results in $\text{GCR}=5/16\approx 31\%$.
        The vertical dotted lines represent frame boundaries, the dashed lines indicate boundaries of the blue segment on the primary channel.
        Frames are not drawn to scale.
        Modified from \cite{vieting2024mixture_encoder}.
    }
    \label{fig:fer}
    \vspace{-1mm}
\end{figure}

%% file: figures/macro_cir_viz.tex
\begin{figure}[tb]
\begin{center}
\begin{tikzpicture}[x=1em, y=6ex, scale=1, line width=0.1em]

    \tikzstyle{mywavstyle}=[yscale=0.6, inner xsep=0, inner ysep=0, fill opacity=0.3, outer sep=0, anchor=west]
    \NewDocumentCommand{\myincludegraphics}{O{} m}{\includegraphics[height=2.8em,#1]{#2}}

    \tikzstyle{mywavstyleTwo}=[yscale=1, inner xsep=0, inner ysep=0, fill opacity=0.3, text opacity=1, outer sep=0, anchor=west]
    \NewDocumentCommand{\myincludegraphicsTwo}{O{} m}{\includegraphics[height=1.68em,#1]{#2}}
    
    \tikzstyle{node}=[draw, circle, inner sep=0.1ex, thin, minimum width=0.5ex, minimum height=0.5ex]
    \tikzstyle{arrow}=[{}-{>}, thin, arrows = {-Stealth[inset=0pt, angle=45:0.4ex]}]
    \tikzstyle{line}=[semithick]
    \tikzstyle{boundary}=[thick, gray, dashed]
    \newcommand{\wordscale}{1};
    \tikzstyle{word}=[
        draw,
        text depth=0pt,
        rectangle,
        text centered,
        rounded corners=0.3ex,
        fill=white,
        semithick,
        inner sep=0pt,
        outer sep=0pt,
        font=\tiny,
        scale=1/\wordscale,
    ] %

    \tikzset{
        lattice/.style={inner sep=0, line width=0, minimum height=0.8em, do path picture={
            \node[node, anchor=west] (wav1node1) at (-1,0) {};
            \node[node, anchor=north] (wav1node2) at (-0.8,1) {};
            \node[node, anchor=north] (wav1node3) at (-0.4,1) {};
            \node[node, anchor=south] (wav1node4) at (-0.6,-1) {};
            \node[node, anchor=south] (wav1node5) at (-0.2,-1) {};
            \node[node, anchor=center] (wav1node6) at (0,0) {};
            \node[node, anchor=south] (wav1node7) at (0.2,-1) {};
            \node[node, anchor=south] (wav1node8) at (0.6,-1) {};
            \node[node, anchor=north] (wav1node9) at (0.4,1) {};
            \node[node, anchor=center] (wav1node10) at (0.8,0) {};
            \node[node, anchor=east] (wav1nodeEnd) at (1,0) {};
            \draw[arrow] (wav1node1) -- (wav1node2);
            \draw[arrow] (wav1node2) -- (wav1node3);
            \draw[arrow] (wav1node3) -- (wav1node6);
            \draw[arrow] (wav1node1) -- (wav1node4);
            \draw[arrow] (wav1node4) -- (wav1node5);
            \draw[arrow] (wav1node5) -- (wav1node6);
            \draw[arrow] (wav1node6) -- (wav1node7);
            \draw[arrow] (wav1node7) -- (wav1node8);
            \draw[arrow] (wav1node6) -- (wav1node9);
            \draw[arrow] (wav1node8) -- (wav1node10);
            \draw[arrow] (wav1node9) -- (wav1node10);
            \draw[arrow] (wav1node10) -- (wav1nodeEnd);
        }},
        latticeSmall/.style={inner sep=0, line width=0, minimum height=0.8em, do path picture={
            \node[node, anchor=west] (wav1node1) at (-1,0) {};
            \node[node, anchor=north] (wav1node2) at (-0.3,1) {};
            \node[node, anchor=south] (wav1node3) at (0.3,-1) {};
            \node[node, anchor=east] (wav1nodeEnd) at (1,0) {};
            \draw[arrow] (wav1node1) -- (wav1node2);
            \draw[arrow] (wav1node1) -- (wav1node3);
            \draw[arrow] (wav1node2) -- (wav1nodeEnd);
            \draw[arrow] (wav1node3) -- (wav1nodeEnd);
        }},
        spoken/.style={inner sep=0, line width=0,  minimum height=0.8em, do path picture={
            \node[minimum height=\y*\wordscale-\pgflinewidth, minimum width=5*\x/20*\wordscale-\pgflinewidth, word, anchor=west] (spokennode1) at ($(-1,0)+(\pgflinewidth,0)$) {Lorem};
            \node[word, minimum height=\y*\wordscale-\pgflinewidth, minimum width=5*\x/20*\wordscale, anchor=west] (spokennode2) at (spokennode1.east) {ipsum};
            \node[word, minimum height=\y*\wordscale-\pgflinewidth, minimum width=4*\x/20*\wordscale, anchor=west] (spokennode3) at (spokennode2.east) {dolor};
            \node[word, minimum height=\y*\wordscale-\pgflinewidth, minimum width=2*\x/20*\wordscale, anchor=west] (spokennode4) at (spokennode3.east) {sit};
            \node[word, minimum height=\y*\wordscale-\pgflinewidth, minimum width=20*\x/100*\wordscale-\pgflinewidth, anchor=west] (spokennode5) at (spokennode4.east) {amet};
        }},
        spokenTwo/.style={inner sep=0, line width=0,  minimum height=0.8em, do path picture={
            \node[minimum height=\y*\wordscale-\pgflinewidth, minimum width=5*\x/20*\wordscale-\pgflinewidth, word, anchor=west] (spokennode1) at ($(-1,0)+(\pgflinewidth,0)$) {wow};
            \node[word, minimum height=\y*\wordscale-\pgflinewidth, minimum width=3*\x/20*\wordscale, anchor=west] (spokennode2) at ($(spokennode1.east)+(6*\x/20*\wordscale,0)$) {let's};
            \node[word, minimum height=\y*\wordscale-\pgflinewidth, minimum width=2.5*\x/20*\wordscale, anchor=west] (spokennode3) at (spokennode2.east) {do};
            \node[word, minimum height=\y*\wordscale-\pgflinewidth, minimum width=2.5*\x/20*\wordscale-\pgflinewidth, anchor=west] (spokennode4) at (spokennode3.east) {it};
        }},
    }

    \node[mywavstyle, fill=muted2, inner sep=0, inner xsep=-0.6em] (wav1) at (0, 0) {\myincludegraphics{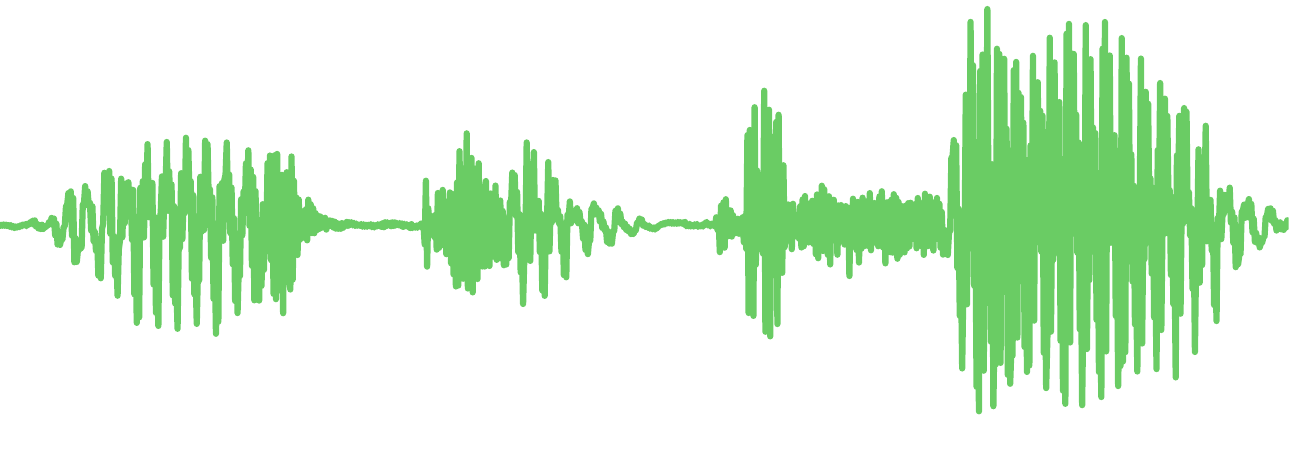}~~~};
    \path (wav1.south west); \pgfgetlastxy{\xsw}{\ysw}; \path (wav1.south east); \pgfgetlastxy{\xse}{\yse};
    \node[lattice,anchor=north west, minimum width=\xse-\xsw] (wav1lat) at ($(wav1.south west) + (0,-0.0)$) {};
    
    \node[mywavstyle, fill=muted3, inner sep=0, inner xsep=-0em, anchor=north] (wav2) at ($(wav1.south west)+(-0.2,-0.6)$) {\myincludegraphics{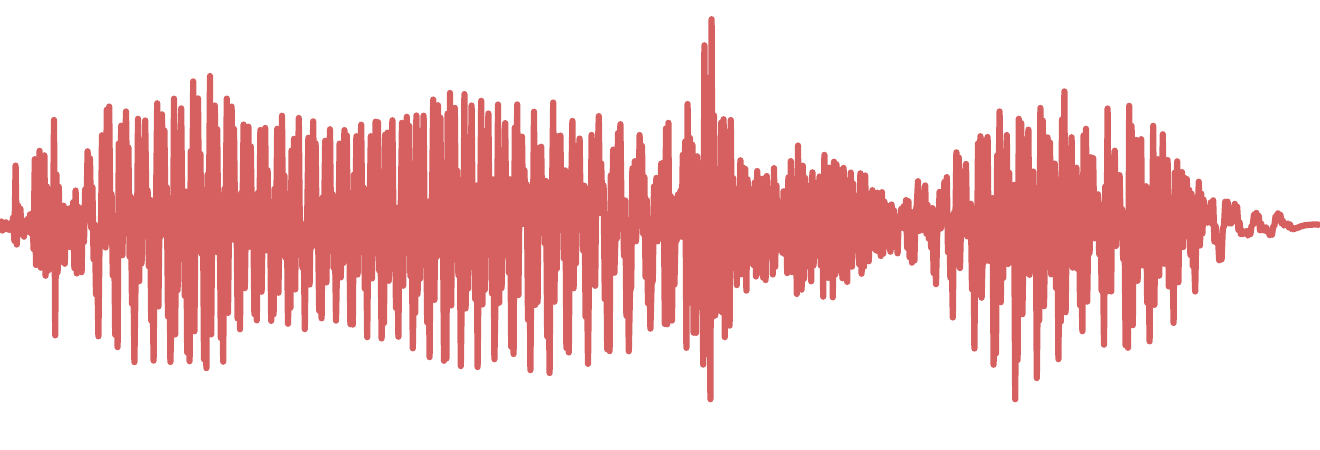}\hspace{0em}};
    \path (wav2.south west); \pgfgetlastxy{\xsw}{\ysw}; \path (wav2.south east); \pgfgetlastxy{\xse}{\yse};
    \node[spoken,anchor=north west, minimum width=\xse-\xsw] (wav2spoken) at ($(wav2.south west) + (0,-0.01)$) {};
    
    \node[mywavstyle, anchor=east, yscale=0.7] (auxSignal) at ($(wav1.east|-wav2)$) {\myincludegraphics[trim=430 0 0 0, clip]{figures/wav/wav_1_C2.pdf}};

    \draw[arrow,{<}-{>}] ($(wav1.north west)+(0,0.25)$) -- node[above=0.25em, outer sep=0, inner sep=0, font=\footnotesize, align=center] {Cross to primary channel\\leakage analysis} ($(wav1.north east)+(0,0.25)$);

    \node[mywavstyleTwo, fill=muted0, inner sep=0,
        anchor=west
    ] (wav3) at ($(wav1.east) + (2.5, 0)$) {%
    \myincludegraphicsTwo[trim=0 -0.12cm 1.5cm 0]{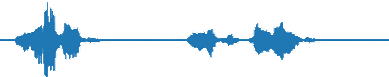}%
    };
    
    \path (wav3.south west); \pgfgetlastxy{\xsw}{\ysw}; \path (wav3.south east); \pgfgetlastxy{\xse}{\yse};
    \node[spokenTwo,anchor=north west, minimum width=\xse-\xsw] (wav3spoken) at ($(wav3.south west) + (0,-0.01)$) {};

    \node[mywavstyleTwo, fill=muted1, inner sep=0, inner xsep=-0.55em,anchor=center] (wav4) at ($(wav3.south east|-wav2)+(-0.2,0)$) {\myincludegraphicsTwo[trim=8cm 0 0 0, clip]{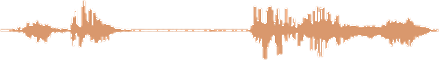}~~~~};

    \path (wav3.south west); \pgfgetlastxy{\xsw}{\ysw}; \path (wav3.south east); \pgfgetlastxy{\xse}{\yse};
    \node[lattice,anchor=north west, minimum width=\xse-\xsw] (wav4lat) at ($(wav4.south west-|wav3.south west) + (0,-0.0)$) {};
    
    \node[mywavstyleTwo, anchor=west, yscale=0.7, inner xsep=0em] (auxSignal) at ($(wav3.west|-wav4)+(0em,0)$) {%
        \myincludegraphicsTwo[trim=0 -0.12cm 10cm 0, clip]{figures/wav/wave_blue.png}%
    };

    \begin{pgfonlayer}{background}
        \draw[boundary] ($(wav1.north west)+(0,0.3)$) -- ($(wav1.north west|-wav2.south)+(0,-0.3)$);
        \draw[boundary] ($(wav1.north east)+(0,0.3)$) -- ($(wav1.north east|-wav2.south)+(0,-0.3)$);
        \draw[boundary] ($(wav3.north west)+(0,0.3)$) -- ($(wav3.north west|-wav4.south)+(0,-0.3)$);
        \draw[boundary] ($(wav3.north east)+(0,0.3)$) -- ($(wav3.north east|-wav4.south)+(0,-0.3)$);
    \end{pgfonlayer}
    \draw[arrow,{<}-{>}] ($(wav3.north west)+(0,0.25)$) -- node[above=0.25em, outer sep=0, inner sep=0, font=\footnotesize, align=center] {Primary to cross channel\\leakage analysis} ($(wav3.north east)+(0,0.25)$);

    \begin{pgfonlayer}{background}
        \draw[line] (wav1-|wav2.west) -- (wav1-|wav4.east);
        \draw[line] (wav2-|wav2.west) -- (wav2-|wav4.east);
    \end{pgfonlayer}

    \coordinate (start) at ($(wav2.south west)+(2.6em,0)$);
    \coordinate (row1) at ($(wav2.south west)+(0,-0.8)$);
    \coordinate (row2) at ($(row1) + (0,-1.2em)$);
    \begin{scope}[local bounding box=legendcol1]
        \node[mywavstyle, anchor=center, yscale=1] (legendWav) at (start|-row1) {\myincludegraphics[trim=199 0 305 0, clip]{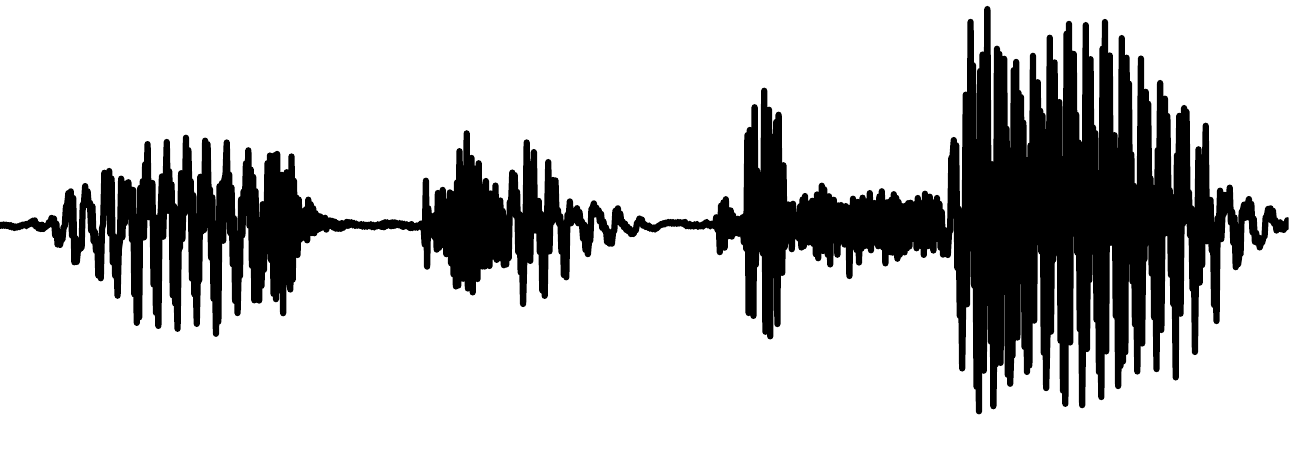}};
        \node[fill=black, fill opacity=0.3, anchor=center, minimum height=1.8ex, minimum width=1.5em] (legendBoundaries) at (start|-row2) {};
    \end{scope}

    \coordinate (start) at ($(legendcol1.east)+(0,0)$);
    \begin{scope}[local bounding box=legendcol2]
        \node[right, font=\footnotesize] (legendWavLabel) at (start|-row1) {waveform};
        \node[right, font=\footnotesize] (legendBoundariesLabel) at (start|-row2) {oracle segment boundaries};
    \end{scope}
    
    \coordinate (start) at ($(legendcol2.east)+(1.5em,0)$);
    \begin{scope}[local bounding box=legendcol3]
        \node[word, minimum height=1em, minimum width=2em, anchor=center] (legendWord) at (start|-row1) {Lorem};
        \node[latticeSmall,anchor=center, minimum width=1.5em] (legendLat) at (start|-row2) {};
    \end{scope}
    
    \coordinate (start) at ($(legendcol3.east)+(0,0)$);
    \node[right, font=\footnotesize] (legendWordLabel) at (start|-row1) {forced alignment};
    \node[right, font=\footnotesize] (legendLatLabel) at (start|-row2) {word lattice};

    \node[fit=(legendWav)(legendWavLabel)(legendBoundaries)(legendBoundariesLabel)(legendWord)(legendWordLabel)(legendLat)(legendLatLabel), inner xsep=0.2em, inner ysep=0em, draw, rounded corners=0.1em, shift={(0,0em), outer sep=0}, thin] () {};

\end{tikzpicture}
\end{center}
    \vspace{-2.0ex}
    \caption{
        Illustration of the leakage analysis from the cross channel to the primary channel (left) and vice versa (right) on the \gls{css} output streams.
        The analysis is illustrated for two (\textcolor{muted2}{green} and \textcolor{peterBlue}{blue}) of the four segments, where in both cases the upper stream is the primary channel and the bottom one the cross channel. For the \textcolor{muted3}{red} and \textcolor{peterOrange}{orange} segment, the roles would be swapped.
    }
    \label{fig:analysis_directions}
    \vspace{-2.0ex}
\end{figure}
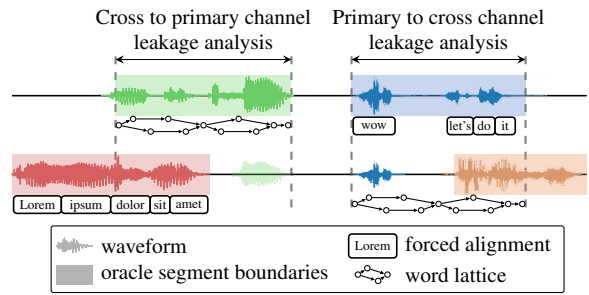

%% file: tables/results_diarizations.tex
\begin{table*}[htbp]

\centering
\begin{tabular}{|c|c|c|c|c|S[table-format=1.1]|}
\hline
                          \multirow{2}{*}{Diarization} &         \multirow{2}{*}{\acs{asr}} &                                                                           \multicolumn{3}{c|}{Data} & {\multirow{2}{*}{\acs{cpwer} [\%]}} \\\cline{3-5}
                                                       &                                    &                              Diarization &             Separation &                       \acs{asr} &     \\\hline\hline
\acs{cssad} (Whisper) \cite{vonneumann2024diarization} &   \multirow{2}{*}{WavLM \acs{aed}} &                                 VoxCeleb & \multirow{2}{*}{LS960} & \multirow{2}{*}{LS960 + Mix94k} & 6.2 \\\cline{1-1}\cline{3-3}\cline{6-6}
                TS-SEP \cite{boeddeker2024interspeech} &                                    &                         LS960 + VoxCeleb &      \multirow{2}{*}{} &               \multirow{2}{*}{} & 6.4 \\\hline
           \multirow{2}{*}{DCF-DS \cite{niu2024dcfds}} &              Transformer \acs{aed} & \multicolumn{2}{c|}{\multirow{2}{*}{VoxCeleb + LS960 + NSF + NF}} &                           LS960 & 6.3 \\\cline{2-2}\cline{5-6}
                                                       &                    WavLM \acs{aed} &                                             \multicolumn{2}{c|}{} &                  LS960 + Mix94k & 4.4 \\\hline\hline
                                                Oracle &              \multirow{3}{*}{Ours} &                                      n/a & \multirow{3}{*}{LS960} &          \multirow{3}{*}{LS960} & 4.3 \\\cline{1-1}\cline{3-3}\cline{6-6}
                                 \acs{cssad} (Whisper) &                                    &                \multirow{2}{*}{VoxCeleb} &                        &                                 & 5.8 \\\cline{1-1}\cline{6-6}
                                    \acs{cssad} (ours) &                                    &                        \multirow{2}{*}{} &      \multirow{2}{*}{} &               \multirow{2}{*}{} & 5.8 \\
\hline
\end{tabular}
\caption{Comparison of \acs{cpwer} on \libricss test with different diarizations in our setup vs. results in the literature. All results use a single microphone. Our results are with the \acs{trafolm}. The other works use \gls{aed} \gls{asr} models. LS960 denotes the LibriSpeech training data, Mix94k refers to the 94k hours WavLM pre-training data. NSF and NF denote the NOTSOFAR and Near Field datasets used for training the front-end in \cite{niu2024dcfds}. VoxCeleb is exclusively used to train the speaker embedding extractors across all works.}
\label{table:results_diarizations}
\vspace{-3.5ex}
\end{table*}

%% file: tables/results_segmentations.tex
\begin{table}[tbp]

\centering
\begin{tabular}{|c|S[table-format=1.1]|S[table-format=1.1]|}
\hline
        \multirow{2}{*}{Segmentation} & \multicolumn{2}{c|}{\acs{orcwer} [\%]} \\\cline{2-3}
                    \multirow{2}{*}{} & {4gram \acs{lm}} &           {Trafo \acs{lm}} \\\hline\hline
                               Oracle &        5.8 &                  4.3 \\\hline
\acs{vad} \cite{vieting2024mixture_encoder} &        7.0 &                  5.6 \\\hline
                \acs{cssad} (Whisper) &        6.7 &                  5.4 \\\hline
                   \acs{cssad} (ours) &        6.5 &                  5.2 \\
\hline
\end{tabular}
\caption{\Acs{orcwer} on \libricss test with different segmentations.}
\label{table:results_segmentations}
\vspace{-2mm}
\end{table}

%% file: tables/results_leakage.tex
\begin{table}[htbp]

\centering
\begin{tabular}{|c|c|c|S[table-format=1.1]|S[table-format=2.1]|S[table-format=2.1]|S[table-format=1.1]|S[table-format=2.1]|}
\hline
                  \multicolumn{2}{|c|}{Hypothesis} &                                  \multicolumn{6}{c|}{Coincidence rate [\%]} \\\hline
\multirow{3}{*}{Primary} &  \multirow{3}{*}{Cross} &  \multicolumn{3}{c|}{Words and sil.} &      \multicolumn{3}{c|}{Words only} \\\cline{3-8}
       \multirow{3}{*}{} &       \multirow{3}{*}{} & \multicolumn{3}{c|}{\#act. speakers} & \multicolumn{3}{c|}{\#act. speakers} \\\cline{3-8}
       \multirow{3}{*}{} &       \multirow{3}{*}{} &  {1} & {2} &                  {Avg.} &  {1} & {2} &                  {Avg.} \\\hline\hline
                  Spoken &                  Spoken &  0.0 & 0.2 &                     0.0 &  0.0 & 0.2 &                     0.0 \\\hline\hline
                  1-best & \multirow{2}{*}{Spoken} &  3.1 & 0.6 &                     2.7 &  0.0 & 0.6 &                     0.1 \\\cline{1-1}\cline{3-8}
                 Lattice &                         &  4.8 & 0.9 &                     4.1 &  0.1 & 0.9 &                     0.2 \\\hline\hline
 \multirow{2}{*}{Spoken} &                  1-best &  8.5 & 0.7 &                     7.1 &  8.2 & 0.7 &                     6.9 \\\cline{2-8}
                         &                 Lattice & 15.2 & 1.3 &                    12.8 & 14.7 & 1.3 &                    12.5 \\
\hline
\end{tabular}
\caption{Analysis of leakage between separated channels. Coincidences are counted once for both words and silence, once for matching words only. The hypotheses (lattices and 1-best) are obtained with the 4gram \gls{lm}. "Spoken" refers to the forced alignment of the ground truth transcription. Results on \libricss test.}
\label{table:results_leakage}

\end{table}

%% file: tables/results_segmentations_fixes.tex
\begin{table}[htbp]

\centering
\begin{tabular}{|c|c|S[table-format=1.1]|S[table-format=1.1]|}
\hline
\multirow{2}{*}{\shortstack{Transcript for\\diarization}} & \multirow{2}{*}{\shortstack{Error types\\eliminated}} & \multicolumn{2}{c|}{\acs{orcwer} [\%]} \\\cline{3-4}
                                        \multirow{2}{*}{} &                                     \multirow{2}{*}{} & {4gram \acs{lm}} &           {Trafo \acs{lm}} \\\hline\hline
                                 \multirow{6}{*}{Whisper} &                                                     - &        6.7 &                  5.4 \\\cline{2-4}
                                        \multirow{6}{*}{} &                                                 Leaks &        6.5 &                  5.2 \\\cline{2-4}
                                        \multirow{6}{*}{} &                                               Missing &        6.1 &                  4.8 \\\cline{2-4}
                                        \multirow{6}{*}{} &                                                Merges &        6.7 &                  5.3 \\\cline{2-4}
                                        \multirow{6}{*}{} &                                            Boundaries &        6.7 &                  5.3 \\\cline{2-4}
                                        \multirow{6}{*}{} &                                                   All &        5.9 &                  4.4 \\\hline\hline
                                    \multirow{6}{*}{Ours} &                                                     - &        6.5 &                  5.2 \\\cline{2-4}
                                        \multirow{6}{*}{} &                                                 Leaks &        6.4 &                  5.2 \\\cline{2-4}
                                        \multirow{6}{*}{} &                                               Missing &        6.0 &                  4.7 \\\cline{2-4}
                                        \multirow{6}{*}{} &                                                Merges &        6.5 &                  5.2 \\\cline{2-4}
                                        \multirow{6}{*}{} &                                            Boundaries &        6.5 &                  5.1 \\\cline{2-4}
                                        \multirow{6}{*}{} &                                                   All &        6.0 &                  4.5 \\
\hline
\end{tabular}
\caption{Effect of different segmentation error types on \gls{orcwer} evaluated based on their elimination using oracle-informed heuristics. Results on \libricss test.}
\label{table:results_segmentations_fixes}
\vspace{-1mm}
\end{table}